\documentstyle[aps,prl,twocolumn]{revtex}
\input{epsf}
\begin{document}
\twocolumn[\hsize\textwidth\columnwidth\hsize\csname@twocolumnfalse\endcsname
\title{Mutator Dynamics on a Smooth Evolutionary Landscape}
\author{David A. Kessler}
\address{Minerva Center and Dept. of Physics, Bar-Ilan University, Ramat Gan, Israel}
\author{Herbert Levine}
\address{Dept. of Physics,
University of California, San Diego La Jolla, CA  92093-0319}
\maketitle
\begin{abstract}
We investigate a model of evolutionary dynamics on a smooth landscape
which features a ``mutator'' allele whose effect is to increase the
mutation rate.  We show that the  expected proportion of mutators
far from equilibrium, when the fitness is steadily increasing in time,
is governed solely by the transition rates into and out of the mutator
state.  This results is a much faster rate of fitness increase than would
be the case without the mutator allele.
Near the fitness equilibrium, however, the mutators are severely
suppressed, due to the detrimental effects of a large mutation rate near
the fitness maximum.  We discuss the results of a  recent experiment on 
natural selection of {\em E. coli} in the light of our model.
\end{abstract}
]
In a recent paper \cite{Snieg}, Sniegowski et. al. presented results concerning the mutation rate of a series of {\em E. coli} populations undergoing natural selection in a laboratory setting.  They found that three out of 
twelve populations had mutated to states with much higher mutation rates, becoming, 
in the language of population genetics, "mutators".  These
changes were traced to the disabling of some specific DNA repair mechanisms 
\cite{miller} in each of the three mutators.  While similar results had been 
obtained in chemostats utilizing artificially hobbled bacterial 
strains \cite{mao,chao}
this is the first time that natural populations had shown selection for more 
rapid mutation.  These results have important implications for a variety of 
topics, including the debate over "directed mutation" \cite{thaler} and the accumulation 
of multiple mutations in cancer cells \cite{cancer}.

Models of mutator selection have in general taken one of two forms.  On one 
hand, Painter \cite{painter} and others \cite{ishii} have considered the case of a
small number of genetic "states" and used mean-field theory to predict mutator
population.  In the simplest realization, a bacterium can be in one of four
states, mutator/non-mutator and high/low fitness.  While this approach may be suitable for the aforementioned
chemostat experiments, it is inadequate to describe the slow overall 
improvement in fitness seen in long-term natural evolution \cite{lenski}.  At the opposite 
extreme, Taddei \cite{taddei} et. al. have produced a complex model by 
combining measured data on {\em E. coli} with other more arbitrary assumptions.
This model, while in principle useful, does not easily lend itself to analysis
and the identification of which details are essential and which not.

In this paper, we propose a simple approach to mutator selection based on
the fitness landscape being smooth.  In previous work \cite{us1,us2}, we and our collaborators
have studied a simple model of evolution in a smooth landscape.  This model
has shown itself very amenable to analytic treatment and in addition capable
of explaining various experimental data on the evolution of RNA 
viruses \cite{rna}.  By
extending the model to include the possibility of a mutator state, we will
show that the basic features of the experimental data can be understood,
provided that the transition rate from the "mutator" state to the normal,
non-mutator state is much slower than the time-scale of the experiment.  The
model then predicts that the system at the time of measurement is in fact
near a crossover between a far-from-equilibrium state where the system is
"climbing" the fitness hill and the equilibrium state; this is consistent
with the measured fitness data \cite{lenski}. The model allows us to make a number of
other predictions which can hopefully be tested by future results of the 
ongoing experiment.

Our basic model assigns a "fitness", i.e. reproduction rate, to each individual
in a population.  This fitness is determined by its genome, which consists of 
$L$ binary genes, with values $0$ and $1$, and is simply the number of $1$'s
in the genome.  Reproduction for each individual is modeled as a Poisson
process whereby a new individual with the same fitness (modulo mutation(
as the parent is added to the population.  At the same time as this
asexual reproduction event occurs, one of the existing members of the 
population is chosen at random and "killed", so as to maintain a fixed 
population size $N$. (The fact that in the experiment growth is allowed to 
take place for some time before the population is culled to its original size,
resulting in a population that oscillates between two values, is an inessential
complication for our considerations.)  Mutation may accompany birth so that,
with probability $\mu$, one of the genes of the baby is chosen at random
and flipped.  The usual assumption in the population genetics literature 
that most mutations are deleterious arises here as a consequence of a 
population lying close to the fitness peak, so that most genes have the 
value $1$.  This creates an imbalance in the overall probability of moving 
up or down the landscape via mutation. The analysis of this model and its 
application to the RNA virus
experiments has been given elsewhere \cite{us1,us2}.

To apply this framework to the {\em E. coli} experiments, we add one special
two-allele gene which controls the mutation rate.  With the mutator allele,
the mutation rate is increased from $\mu$ to $\lambda\mu$ without any direct
effect on fitness.  We take $\sigma_f$, ($\sigma_b$) to be the forward 
(backward) mutation rate to (from) the mutator state. 

We first analyze the equilibrium state of the population.
Assuming $N \gg L$, we can use mean-field theory and ignore any fluctuations in
the population.  This leads immediately to the equations
\begin{eqnarray}
0 = \dot P_x &=& (x-\bar x ) P_x + \mu x ( p_x P_{x-1} +(1-p_x) P_{x+1}-
P_x) \nonumber \\
&\ & -\ \sigma_f xP_x + \sigma_b x Q_x \nonumber \\
0 = \dot Q_x &=& (x - \bar x ) Q_x + \lambda \mu x (p_x Q_{x-1} +(1-p_x) Q_{x+1}
-Q_x) \nonumber \\ 
&\ & +\ \sigma_f x P_x - \sigma_b x Q_x 
\end{eqnarray}
Here, $p_x$ is the probability that a mutation results in moving up in 
fitness, which equals $1-x/L$. $P_x$ and $Q_x$ are respectively the normal 
and mutator population fractions at fitness $x$, and $\bar x$ is the mean
fitness of the population, considering both normal and mutator types on an
equal footing.  To solve this equation, we assume that $L$ is large and
define $y\equiv L-x$.  To leading order in $L$, 
\begin{eqnarray}
0 &=& (\bar y - y) P_y + \mu L (-P_y+P_{y-1}) - \sigma_f LP_y
+ \sigma_b L Q_y  \\
0 &=& (\bar y - y) Q_y + \lambda \mu L (-Q_y+Q_{y-1}) + 
\sigma_f L P_y - \sigma_b L Q_y 
\end{eqnarray}
This equation also applies for $y=0$ if we set $P_{-1}=Q_{-1}=0$.  At $y=0$, 
we obtain
\begin{equation}
Q_0= P_0 \frac{\sigma_f L}{\sigma_b L + \lambda\mu L - \bar y}
\end{equation}
as well as a quadratic equation for $\bar y$.  While this equation is messy,
it is easy to check that if $\lambda=1$, then $\bar y=\mu L$, giving
$\sigma_b Q_0 = \sigma_f P_0$ as expected by naive balance between the
forward and backward transitions.  On the other hand, if $\lambda$ is large, 
$\bar y=(\mu+\sigma_f)L$; this is easily seen since $Q_0 \approx (\sigma_f/\lambda\mu)P_0$ is then small
and can be dropped from the $P_0$ equation.  

To make further progress, we let $\lambda$ be large.  Then, to leading order, 
the $P$ subpopulation decouples and 
\begin{equation}
P_y = P_0 \frac{(\mu L)^y}{\Gamma(y + 1)}
\end{equation}
where $P_0$ is fixed by normalization to $P_0=e^{-\mu L}$.  This is precisely
the result in the absence of the mutator state.  The $Q$ distribution 
follows from the inhomogeneous recursion relation, eq. (2), with the
sources fixed by the known $P_y$.  Solving this equation, we see that
the relative narrowness of the $P$ distribution essentially ``collapses''
all the source terms to $y=0$, giving the simple result 
for $1 \ll y \ll \lambda$, 
\begin{equation}
Q_y \sim \frac{\sigma_f}{\lambda \mu } \left[\frac{(\lambda\mu L)^y \Gamma(\lambda \mu L + 1)}{\Gamma(\lambda \mu L + y + 1)}\right] \sim \frac{\sigma_f}{\lambda \mu } e^{-y^2/(2\lambda\mu L)}
\end{equation}
Thus, even though each individual $Q_y$ is of order $\sigma_f/\lambda\mu$,
the width of the $q$ distribution is large, of order $\sqrt{\lambda
\mu L}$, so the total number of fast individuals is much larger,
$\sigma_f \sqrt{{{2 L} \over {\pi \lambda \mu }}}$.
One can further check this calculation by computing the value of $\bar y$ from
the $P$ and $Q$ distributions. The mean $y$ of the normal type is just
$\mu L$, but the small number of fast type have anomalously high $y$ and
contribute a total of $\sigma_f L$ to the total fitness without changing
the total number of individuals significantly, reproducing the result obtained
above.

The reason for the suppression of the fast type is clear, and has been
appreciated for a long time \cite{leigh}.  The ``natural'' equilibrium of the
fast type, if there were no slow type around, would place them $\lambda$
times as far from the fitness peak than the slow type.  Thus, the faster
mutation rate of the fast species places a ``mutational load'' 
on them, so they lose out in competition with the normal species and are
suppressed.  The noteworthy part of the calculation is the prediction
size of the suppression, which would be difficult to guess {\em a priori}.

We now turn to the far-from-equilibrium case where the typical fitness of an
individual is close to $L/2$ on the scale of $L$.  Here, mutations will produce movement up or down the landscape with equal probability.  In this case,
with just the normal type present, we found \cite{us2} that the population increases 
its mean fitness at a constant rate on average.  In this regime, mean-field 
theory is an utter failure, leading \cite{us1,us2} to finite-time singularities caused by
the inability to properly account for the essential variance limiting 
property of birth-death processes \cite{zhang}.  We turn instead to an approach \cite{us2} which 
involves truncating the state space of the exact Markov process, utilizing 
the assumption that the mutation probabilities $\mu$, $\sigma_f$ and 
$\sigma_b$ are much smaller than unity.  To see the structure of the
problem, we focus first on the case where the population consists of only
two individuals, i.e. $N=2$.  In this case, the states we need to consider
are those with the two individuals having the same fitness $x$, either
both normal, $(f11)_x$, both mutator, $(f22)_x$, or one normal and one
mutator, $(f12)_x$.  In addition, we need to consider the states where
the two individuals have adjacent fitnesses at $x$ and $x+1$, with both
being either normal, $(g11)_x$, or mutator, $(g22)_x$.  The states $(g12)_x$
and $(g21)_x$ where the two individuals having different fitness and type
are of order $\mu\sigma_f$ and are therefore dropped along with the states
where the two individuals differ in fitness by more than 1, which are higher
order in $\mu$.  Truncating the master equation to these states yields
\begin{eqnarray}
\dot{(f11)}_x &=& -2\mu x(f11)_x + \frac{1}{2}x(g11)_x 
  + \frac{1}{2}x(g11)_{x-1} \nonumber \\
&\ & -\ 4\sigma_f x (f11)_x + \frac{1}{2}x(f12)_x
\end{eqnarray}
and similar equations for the other 4 densities.
In the case with just normal individuals, we found \cite{us2} that the 
long-time solution had the scaling form $f_x = {1 \over x} F(x/t)$, which 
indicates directly
the linear growth of fitness with time.  Examining our system of equations,
we see that such a scaling form is not possible unless
\begin{eqnarray}
\sigma_b f(22) &=& \sigma_f f(11) + (\Delta 22)(x/t)/(x t^2) \\
f(12) &=& 8\sigma_f f(11) + (\Delta 12)(x/t)/(x t^2)
\end{eqnarray}
The physical meaning of these relationships is that the system must quickly
evolve to a state of "local equilibrium" between the normal and fast 
types before
it can achieve the linear-velocity climbing state.  In this ``local 
equilibrium'', the ratio of fast to normal types is given by 
$2(f22) + f12)/2((f11)+f12+f22)=\sigma_f/\sigma_b$, just
as it would if there were no fitness degrees of freedom.  Substituting this
scaling form into the truncated master equations and dropping all time 
derivatives of the $g$'s
and of $f12$ (they are second order quantities), we get, after eliminating
the $\Delta$ terms
\begin{equation}
(1 + \frac{\sigma_f}{\sigma_b})\dot{f11} = (1 + \frac{\lambda\sigma_f}{\sigma_b})\left\{ \frac{\mu}{2}\left[ (f11)' + x(f11)''\right] \right\}
\end{equation}
This equation shows that the velocity has been renormalized by the factor
$(\sigma_f \lambda + \sigma_b)/(\sigma_f + \sigma_b)$, which is very reasonable
given the ratio of mutator to normal types and the fact that the velocity 
is exactly linear in the mutation rate and the number of
individuals. 

The above methodology can be extended to arbitrarily large $N$ with the
basic results concerning the percentage of mutators and the renormalization
of the velocity unchanged.  Basically, they follow immediately from the above
scaling structure of the long-time solution.  In Fig. 1,
we show the prediction of this analysis along with direct 
simulations of our Markov process. The agreement of the theory with
the asymptotic state is quite satisfactory. There is however a transient period
in which a larger number of mutators are selected. This latter behavior is due 
to the fact
that evolutionary advance of the {\em initial} population (consisting entirely of
wild-type cells) occurs most readily by creating mutators; eventually, these
mutators start making back transitions and establish the steady-state ratio. This
transient response is analogous to what has been termed "hitch-hiking" in the
genetics literature. We see that this naive picture of the selection of mutators
in the climbing state breaks down rather quickly and the true situation
is one of balance based solely on the forward and backward rates.

In a given experiment, the population is placed in a new environment
and allowed to evolve towards the fitness peak.  Initially, the rate
of fitness improvement will be correlated with the mutator percentage
in any given realization, with the average effect as given above.  Eventually, 
the system will begin to approach equilibrium and the mutators will start
to be at a disadvantage.  Roughly, the crossover should take place when the average fitness
equals the equilibrium mean fitness of the mutator type considered alone,
namely a distance $\lambda \mu L$ from the peak.  After this point, it is advantageous for the population to shed its mutators and continue to climb
to its final resting point, a distance $(\mu + \sigma_f)L$ from the top.
In essence, then, the
system gets a "free-ride" from the presence of the mutators. They help the
system to get close to the fitness peak, but then politely bow out of the
picture so as not to significantly impair the final mean fitness.

\begin{figure}
\centerline{\epsfxsize=3.25in \epsffile{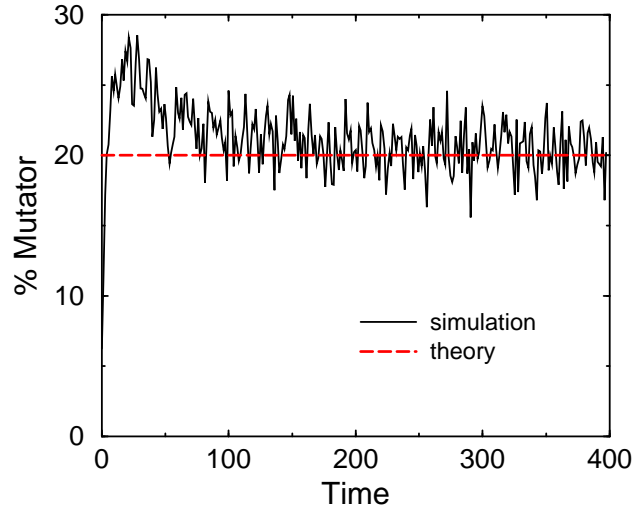}}
\caption{Percentage of mutators in the far-from-equilibrium case, with
$\sigma_f=0.001$, $\sigma_b=0.004$, together with the asymptotic
prediction. $\mu=0.01$, $\lambda=10$, $N=100$, averaged over 400 
realizations.}
\end{figure}

We now turn to a discussion of the results of the experiment detailed in
Ref. \cite{Snieg}.  This work reported that three out of twelve populations fixed completely the mutator genotype; yet there was no clear correlation between fitness and mutator status at the current epoch.  While
it is possible that fluctuations are dominating the average behavior,
we will assume the experimental results reflect
the typical behavior of the system and ask how they relate to our model.
It is clear that the only way to make our model correspond to the
experimental findings is to take the backward mutation rate $\sigma_b$ to
be so small as to play no role on the time-scale of the experiment. 
Otherwise, one would have expected to see reversions from the mutator
state to the normal state, which were not observed.  
Biologically, this vanishingly small $\sigma_b$ corresponds to the mutation 
of the DNA repair mechanism being due to something other than a (reversible)
point mutation.  This assumption of no back-mutations implies that it is
possible to completely fix the mutator gene, the
percentage of systems doing so increasing monotonically with time.
The rate of fixation becomes vanishingly small as the mean fitness in the 
non-fixed systems approaches the equilibrium mutator fitness.  After
this crossover point, the mutators become suppressed and fixation becomes extremely
unlikely.  This behavior is indicated in Fig. 2, in which a set of 40
populations experienced nine instances of mutator fixation.  The fact that
no additional fixations were seen in the last third
of the experiment indicates that the system was near this crossover point
at the end of the run.  This fact is consistent with the leveling off of the
rate of fitness improvement in the experiment, as report in Ref. \cite{lenski} .  Near
the crossover point, there is no clear correlation between mutator status
and fitness, as can be verified from  our simulations (see Fig. 3).  
This lack of correlation is also consistent with the experimental
findings.

\begin{figure}
\centerline{\epsfxsize=3.25in \epsffile{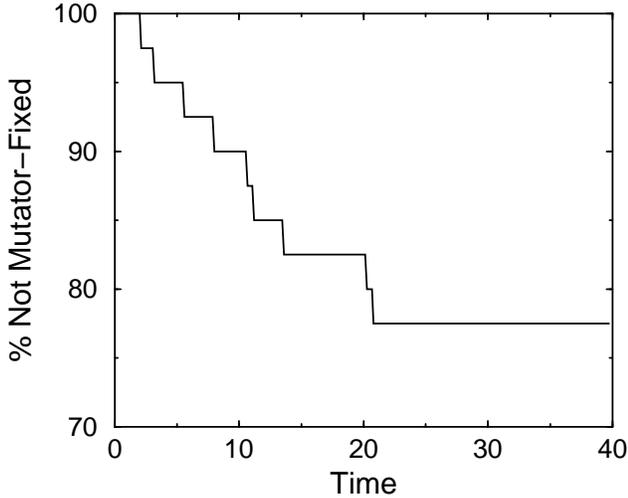}}
\caption{Fraction of populations which have not completely fixed the mutator
gene, for a simulation with 40 independent realizations of our stochastic evolution model.
$\mu = .005$, $\sigma _f =.0005$, $\sigma_b=0$, $\lambda = 15$,
$L  = 100$, $N=1000$. After the times
shown, no more populations became mutator-fixed.}
\end{figure}

If our understanding proves correct, we can make the following predictions
for what should be seen as the experiment is continued past the times
reported.
Since $\sigma_b$ is small, no reversions should be seen.  Since we are near the
crossover, we expect no additional mutator fixations. More significantly,
the correlation between fitness and mutator status should have been positive
in the past and should become negative in the future.  This can be tested by studying isolates from fixed times in the past and by repeating the measurements
as the evolution continues.  Also,  we predict that the percentage of mutation
fixations should depend on the initial distance from the fitness peak; that is,
how much the experimental growth conditions depart from the usual wild-type
habitat.

To summarize, we have presented a simple model which describes how the 
dynamics of mutation rate changes affects and is affected by natural
selection.  The analytic tractability of our evolutionary dynamics allows us to
focus directly on the issues which determine selection for or against mutator
states.  Our various predictions for the experiment should test
whether this type of generic analysis is powerful enough to make reliable statements regarding these complex biological processes.

\begin{figure}
\centerline{\epsfxsize=3.25in \epsffile{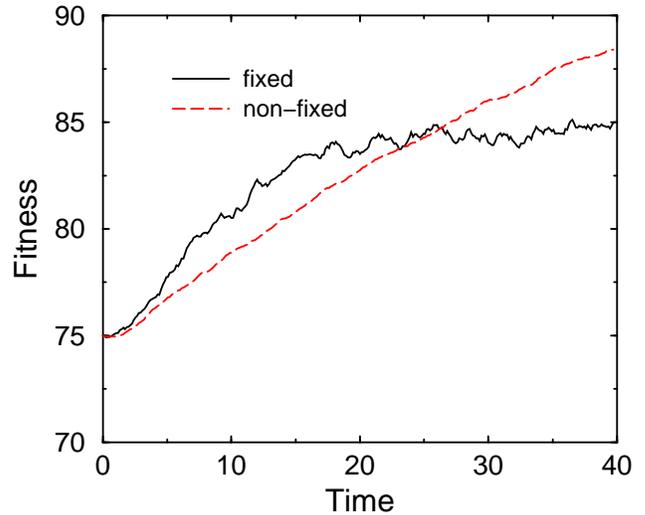}}
\caption{Fitness vs. time, for the simulations described in
Figure 2, plotted separately the average fitness over the subset of our
populations which fix (9 out of 40) or do not fix (31 out of 40) the
mutator gene sometime during the run. }
\end{figure}

HL acknowledges the support of the US NSF under grant DMR94-15460; DAK acknowledges
the support of the Israel Science Foundation.

\end{document}